\documentclass[prd,reprint,twoside,amsmath,nofootinbib,showpacs]{revtex4-1}

\usepackage[T1]{fontenc}
\usepackage[utf8]{inputenc}
\usepackage[brazil,english]{babel}
\usepackage{tensor}
\usepackage{cancel}
\usepackage[usenames,dvipsnames]{xcolor}
  \definecolor{ultramarine}{RGB}{63,0,255}
\usepackage{amssymb} 
\usepackage{mathtools}
\usepackage{graphicx}
\usepackage{nicefrac}
\usepackage{paralist}
\usepackage{fancyhdr}
\makeatletter
  \fancypagestyle{plain}{\fancyhead{}}
\makeatother
\usepackage[unicode,pdfusetitle]{hyperref}
\hypersetup{
  pdfinfo = {
    Author = {Henrique A. Gomes, Daniel C. Guariento},
    Title = {Hamiltonian analysis of the cuscuton}
  },
  colorlinks = true,
  breaklinks = true,
  linkcolor = Blue,
  urlcolor = Blue,
  citecolor = Blue
}
\usepackage[textsize=tiny,linecolor=orange,backgroundcolor=orange!40,colorinlistoftodos]{todonotes}

\makeatletter
\providecommand*{\diff}%
{\@ifnextchar^{\DIfF}{\DIfF^{}}}
\def\DIfF^#1{%
  \mathop{\mathrm{\mathstrut d}}%
  \nolimits^{#1}\gobblespace}
\def\gobblespace{%
  \futurelet\diffarg\opspace}
\def\opspace{%
  \let\DiffSpace\!%
  \ifx\diffarg(%
  \let\DiffSpace\relax
  \else
  \ifx\diffarg[%
  \let\DiffSpace\relax
  \else
  \ifx\diffarg\{%
  \let\DiffSpace\relax
  \fi\fi\fi\DiffSpace}

\providecommand*{\deriv}[3][]{%
  \frac{\diff^{#1}#2}{\diff #3^{#1}}}

\providecommand*{\fderiv}[3][]{%
  \frac{\delta^{#1}#2}{\delta #3^{#1}}}

\DeclareMathOperator{\Lie}{\mathcal{L}}

\allowdisplaybreaks[1]

\begin{document}

\title{Hamiltonian analysis of the cuscuton}

\author{Henrique Gomes}
\email{hgomes@perimeterinstitute.ca}

\affiliation{Perimeter Institute for Theoretical Physics, 31 Caroline St.\ N., Waterloo, ON, N2L 2Y5, Canada}

\author{Daniel C.\ Guariento}
\email{dguariento@perimeterinstitute.ca}

\affiliation{Perimeter Institute for Theoretical Physics, 31 Caroline St.\ N., Waterloo, ON, N2L 2Y5, Canada}

\begin{abstract}

The cuscuton was introduced in the context of cosmology as a field with infinite speed of propagation. It has been claimed to resemble Ho\v{r}ava gravity in a certain limit, and it is a good candidate for an ether theory in which a time-dependent cosmological constant appears naturally. The analysis of its properties is usually performed in the Lagrangian framework, which makes issues like the counting of its dynamical degrees of freedom less clear-cut. Here we perform a Hamiltonian analysis of the theory. We show that the homogeneous limit with local degrees of freedom has singular behavior in the Hamiltonian framework. In other frames, it has an extra scalar degree of freedom. The homogeneous field has regular behavior only if defined \emph{a priori} as a spatially constant field in a CMC foliation and contributing with a single global degree of freedom. Lastly, we find conditions on the cuscuton potential for the resulting lapse function to be non-zero throughout evolution.
  
\end{abstract}


\maketitle

\section{Introduction}

One of the main observational issues facing cosmology and gravitational theory are the phases of accelerated expansion of the Universe: the many forms of the cosmological constant problem, dark energy, inflation. Moreover, given the failure of particle physics to account for the observed value of the vacuum energy, a quantum gravity theory is expected to at least give some insight on a solution to this problem. There have been several attempts at tackling one form or another of dark energy properties, mostly concentrating on two approaches: modifying general relativity or considering an additional field that would be responsible for these large-scale differences while retaining the well observed short-distance predictions of general relativity.

There is a considerable overlap between these approaches, as many modified gravity theories admit a description in the so-called ``Einstein frame'', in which they are formulated as general relativity coupled to a scalar field. In fact, an economic and elegant way for scalar-tensor theories of gravity to take advantage of this fact when interpreting modifications of gravity is to adopt the effective description of a time-dependent cosmological constant and to provide constraints on this variability \cite{Afshordi:2015iza}. When formulating this system, the most general scalar-tensor theory of gravity that generates only second-order equations of motion is the Horndeski theory \cite{Horndeski:1974wa}, also referred to in the literature as the generalized Galileons \cite{Deffayet:2011gz,*Zumalacarregui:2012us,*Gao:2011qe}. There are candidates to generalizing this theory to include higher-order spatial derivatives, known as beyond-Horndeski theories \cite{Gleyzes:2014dya,*Gleyzes:2014qga}.

A simple yet noteworthy subset of Horndeski theories is the extreme form of $k$-essence known as the \emph{cuscuton} \cite{Afshordi:2006ad,*Afshordi:2009tt,*Afshordi:2007yx}. It is defined as the limit of $k$-essence in which the sound speed, as defined by derivatives of the Lagrangian with respect to the kinetic term of the scalar field, becomes infinite \cite{vikman-2007}. Interestingly, in flat or cosmological backgrounds, as well as in some other special cases \cite{Abdalla:2013ara}, the  equations of motion of the cuscuton are degenerate, effectively becoming only a source term in the Hamiltonian constraint of general relativity. Its behavior is akin to a time-dependent cosmological constant, without any propagating excitations. This degenerate behavior in the homogeneous limit has prompted the characterization of the cuscuton as a field that adds no additional degrees of freedom to a gravitational system, and, because the homogeneity condition selects a preferred foliation, the cuscuton has since been identified with similarly degenerate limits of modified gravity theories with broken Lorentz invariance, such as Ho\v{r}ava--Lifshitz gravity \cite{Afshordi:2009tt}. Correspondence with Einstein--\ae{}ther theory is still a subject of debate \cite{Bhattacharyya:2016mah}.

The cuscuton is in fact not the unique subclass of Horndeski that possesses such a degenerate behavior. It was recently shown that a more general subclass of fields, known as \emph{kinetic gravity braiding} \cite{Deffayet:2010qz,*Pujolas:2011he}, also produces degenerate equations of motion under certain limits, and a class of exact solutions of general relativity has been found with this field as source \cite{Afshordi:2014qaa}.

All these developments motivate one to ask: does the cuscuton in fact possess no propagating degrees of freedom? If this is generically true, a theory of cosmological perturbations need not worry about infinite propagation speeds, since no excitations would travel in general. Conversely, if there are in fact cases in which the cuscuton does propagate degrees of freedom, one needs to be aware of whether these cases pose a problem for causality and an incompatibility with Lorentz invariance in general. 

The purpose of this work is to address this question. To do that in a rigorous setting, we formulate the cuscuton system coupled to general relativity formally using the Hamiltonian formalism and infer its dynamics by using the algebra of constraints to count the degrees of freedom.

The paper is organized as follows: in Sec.\ \ref{sec:cusc-ham} we state the generic inhomogeneous cuscuton system coupled to general relativity in the Hamiltonian language. Section \ref{sec:symplectic} discusses the symplectic geometry of this system and the constraint algebra which leads to the counting of degrees of freedom. The homogeneous limit and its pathologies are explored in Sec.\ \ref{sec:homog-strong}. We present our conclusions and a brief discussion of the results in Sec.\ \ref{sec:conclusion}. Throughout this work, Latin indices run from 1 to 3, an overhead dot represent time derivatives, and we use a mostly-plus Lorentzian signature.

\section{Hamiltonian cuscuton dynamics}\label{sec:cusc-ham}

The cuscuton Lagrangian \cite{Afshordi:2006ad,*Afshordi:2009tt,*Afshordi:2007yx} is a particular case of $k$-essence \cite{ArmendarizPicon:2000dh,*ArmendarizPicon:2000ah}, in its turn defined as the subset of the Horndeski \cite{Horndeski:1974wa} action with only $L_2 \neq 0$. Its scalar sector is given in the language of generalized Galileons \cite{Gleyzes:2014dya,*Gleyzes:2014qga} by
\begin{equation}\label{cuscuton-lagrangian}
  L_2 = \int \diff^3 x \, \sqrt{-g} \left[ \mu^2 \sqrt{\left| 2 X \right|} - V(\phi) \right]\,,
\end{equation}
with $g$ the determinant of the spacetime metric, and with $\mu$ a coupling constant. The kinetic term $X$ is defined in terms of the first derivatives of $\phi$ as
\begin{equation}\label{Xdef}
  X = \frac{1}{2} \left[ \frac{1}{N^2} \left( \dot{\phi} - \xi^i \nabla_i \phi \right)^2 - \gamma^{i j} \nabla_i \phi \nabla_j \phi \right]\,,
\end{equation}
where we have performed the ADM decomposition of the spacetime metric in terms of the lapse $N$, shift $\xi^i$ and spatial metric $\gamma_{i j}$, so that $\sqrt{-g} = N \sqrt{\gamma}$. From here on, we use the metric $\gamma$ to raise and lower indices as per the ADM notation \cite{adm-1962}. Also note that the modulus in Eq.\ \eqref{cuscuton-lagrangian}, while ensuring that the Lagrangian be always real, also defines two branches depending on the sign of the kinetic term $X$. Since the main interest of the cuscuton are its cosmological applications, the interesting region should be the one in which the homogeneous field constitutes a spacelike surface. Therefore, for the remainder of this work, we consider only the branch $X > 0$.\footnote{See Appendix \ref{sec:Xneg} for comments on the negative branch.}

The canonical momentum associated with $\phi$ is 
\begin{equation}\label{eq:pidephi}
  \begin{split}
    \pi_\phi \equiv&\, \fderiv{L_2}{\dot{\phi}} \\
    =&\, \frac{\sqrt{\gamma}}{N} \frac{\mu^2}{\sqrt{2 X}} \left( \dot{\phi} - \xi^i \nabla_i \phi \right) \,,
  \end{split}
\end{equation}
and, in general, this definition does not introduce any primary constraints. The expression \eqref{eq:pidephi} may be solved for $\dot{\phi}$, giving
\begin{equation}\label{eq:phidepi}
  \dot{\phi} =  \xi^i \nabla_i \phi \pm N \pi_\phi \sqrt{\frac{\nabla^i \phi \nabla_i \phi}{\pi_\phi^2 - \gamma \mu^4}} \,.
\end{equation}
One should notice, however, that the solution with the minus sign above is spurious. If one inserts the negative branch of Eq.\ \eqref{eq:phidepi} into Eq.\ \eqref{eq:pidephi}, one obtains $\pi_\phi = -\pi_\phi$. Therefore, the only reasonable solution is the one with the plus sign. Expressing the kinetic term $X$ in terms of $\pi_\phi$ as
\begin{equation}
  X = \frac{1}{2} \nabla^i \phi \nabla_i \phi \left( \frac{\gamma \mu^4}{\pi_\phi^2 - \gamma \mu^4} \right) \,,
\end{equation}
the Legendre transform will yield the following cuscuton Hamiltonian
\begin{equation}\label{eq:cuscuton-H}
  \begin{split}
    H_\phi =&\, \int \diff^ 3x\, \pi_\phi \dot{\phi} - L_2 \\
    =&\, \int \diff^ 3x\,\left[ N \sqrt{\nabla^i \phi \nabla_i \phi \left( \pi_\phi^2 - \gamma \mu^4 \right)} + N \sqrt{\gamma} V \right.\\
    &\left. \vphantom{\sqrt{\nabla^i \phi \nabla_i \phi \left( \pi_\phi^2 - \gamma \mu^4 \right)}} + \xi_i \pi_\phi \nabla^i \phi \right]\,.
  \end{split}
\end{equation}

For ease of comparison with the corresponding terms that come from the gravitational sector when computing the Hamiltonian and momentum constraints, the Hamiltonian \eqref{eq:cuscuton-H} may be cast as a sum of a superhamiltonian $\mathcal{H}_\phi$ and a supermomentum $\mathcal{H}^i_\phi$, that is,
\begin{equation}
  H_\phi = \int \diff^3 x \left( N \mathcal{H}_\phi + \xi_i \mathcal{H}^i_\phi \right) \,,
\end{equation}
where
\begin{align}
  \mathcal{H}_\phi =&\, \sqrt{\nabla^i \phi \nabla_i \phi \left( \pi_\phi^2 - \gamma \mu^4 \right)} + \sqrt{\gamma} V\,, \label{ham-constraint-cusc}\\
  \mathcal{H}^i_\phi =&\, \pi_\phi \nabla^i \phi \,.
\end{align}
Ignoring surface terms (by assuming that the spatial manifold is compact without boundary), the equations of motion of this field then read\footnote{See Appendix \ref{sec:full-var} for the derivation of the full variation.}
\begin{align}
  \begin{split}
    \dot{\pi}_\phi \equiv&\, \{\pi_\phi, H\}= - \fderiv{H_\phi}{\phi}\\
    =&\, \nabla_i \left( N \nabla^i \phi \sqrt{\frac{\pi_\phi^2 - \gamma \mu^4}{\nabla^a \phi \nabla_a \phi}} + \xi^i \pi_\phi \right) \!- N \sqrt{\gamma} \deriv{V}{\phi} ,
\label{eq:dot_pi_phi}  \end{split}\\
  \begin{split} \label{eq:dot-phi}
    \dot{\phi} \equiv&\, \{ \phi, H \} = \fderiv{H_\phi}{\pi_\phi}\\
    =&\, \xi^i \nabla_i \phi + N \pi_\phi \sqrt{\frac{\nabla^i \phi \nabla_i \phi}{\pi_\phi^2 - \gamma \mu^4}} \,.
  \end{split}
\end{align}

The gravitational sector of the Hamiltonian in the ADM decomposition \cite{adm-1962} reads
\begin{equation}
  \begin{split}
    H_\text{g} =&\, \int \diff^3 x \left\{ N \left[ \frac{1}{\sqrt{\gamma}} \left( \pi^{a b} \pi_{a b} - \frac{1}{2} \pi^2 \right) - \sqrt{\gamma} \mathcal{R} \right] \right.\\
    &\left. \vphantom{\frac{1}{\sqrt{\gamma}}} + 2 \xi_a \nabla_b \pi^{a b} \right\} \\
    =&\, \int \diff^3 x \left( N \mathcal{H}_{\text{g}} + \xi_a \mathcal{H}_{\text{g}}^a \right) \,,
  \end{split}
\end{equation}
where $\mathcal{R}$ is the 3-Ricci scalar and we have defined
\begin{align}
  \mathcal{H}_{\text{g}} \equiv&\, \frac{1}{\sqrt{\gamma}} \left( \pi^{a b} \pi_{a b} - \frac{1}{2} \pi^2 \right) - \sqrt{\gamma} \mathcal{R} \,, \label{ham-constraint-g}\\
  \mathcal{H}_{\text{g}}^a \equiv&\, 2 \nabla_b \pi^{a b} \,,
\end{align}
so the full Hamiltonian of the cuscuton field minimally coupled to the gravitational field reads
\begin{equation}\label{eq:action-full}
  \begin{split}
    H =&\, H_{\text{g}} + H_\phi\\
    =&\, \int \diff^3 x \left[ N \left( \mathcal{H}_{\text{g}} + \mathcal{H}_\phi \right) + \xi_a \left( \mathcal{H}_{\text{g}}^a + \mathcal{H}_\phi^a \right) \right] \,.
  \end{split}
\end{equation}
Clearly, the Hamiltonian has cyclic variables, imposing $\pi_N = 0 = \pi_\xi$. With the full action in mind, we may now consider the Hamiltonian and momentum constraints. Computing the Poisson brackets of the Hamiltonian \eqref{eq:action-full} with the lapse and shift momenta, we find
\begin{align}
  \begin{split}\label{Ham_constraint}
    \left\{ H, \pi_N \right\} =&\, \frac{1}{\sqrt{\gamma}} \left( \pi^{a b} \pi_{a b} - \frac{1}{2} \pi^2 \right) - \sqrt{\gamma} \mathcal{R} \\
    & + \sqrt{\nabla^i \phi \nabla_i \phi \left( \pi_\phi^2 - \gamma \mu^4 \right)} + \sqrt{\gamma} V \\
    =&\, 0 \,,
  \end{split}\\
  \left\{ H, \pi_\xi^a \right\} =&\, 2 \nabla_b \pi^{a b} + \pi_\phi \nabla^a \phi = 0 \,,\label{diff_constraint}
\end{align}
and the Poisson brackets of the full Hamiltonian \eqref{eq:action-full} with the fields yield the equations of motion \cite{adm-1962}. Together with Eqs.\ \eqref{eq:dot_pi_phi} and \eqref{eq:dot-phi}, the remaining equations of motion are
\begin{align}
  \begin{split}\label{eq:dot-gamma}
    \dot{\gamma}_{i j} \equiv&\, \left\{ \gamma_{i j}, H \right\}\\
    =&\, 2 \frac{N}{\sqrt{\gamma}} \left( \pi_{i j} - \frac{1}{2} \pi \gamma_{i j} \right) + 2 \nabla_{\left( i \right.} \xi_{\left. j \right)} \,,
  \end{split}\\
  \begin{split}
    \dot{\pi}^{i j} \equiv&\, \left\{ \pi^{i j}, H \right\}\\
    =&\, - N \sqrt{\gamma} \left( \mathcal{R}^{i j} - \frac{1}{2} \gamma^{i j} \mathcal{R} \right)\\
    & + \frac{N}{\sqrt{\gamma}} \left[ \frac{\gamma^{i j}}{2} \left( \pi^{c d} \pi_{c d} - \frac{1}{2} \pi^2 \right) \right.\\
    & \left. - 2 \left( \pi^{i c} \, \tensor{\pi}{_c ^j} - \frac{1}{2} \pi \, \pi^{i j} \right) \right] \\
    & + \sqrt{\gamma} \left( \nabla^i \nabla^j N - \gamma^{i j} \nabla^c \nabla_c N \right) \\
    & + \nabla_c \left( \pi^{i j} \xi^c \right) - 2 \pi^{c \left( i \right.} \nabla_c \xi^{\left. j \right)} \\
    & + \frac{N}{2} \left[ \nabla^i \phi \nabla^j \phi \sqrt{\frac{\pi_\phi^2 - \gamma \mu^4}{\nabla^a \phi \nabla_a \phi}} \right.\\
    & \left. + \gamma^{i j} \left( \gamma \mu^4 \sqrt{\frac{\nabla^a \phi \nabla_a \phi}{\pi_\phi^2 - \gamma \mu^4}} - V \sqrt{\gamma} \right) \right] \,.
  \end{split}
\end{align}

Now, the constraints \eqref{Ham_constraint} and \eqref{diff_constraint} might have first and second class components. A non-trivial second class part might imply that refoliation invariance be partly gauge-fixed. In that case, the theory would come equipped with a ``preferred frame''. In order to ascertain whether this is the case, we now analyze the symplectic geometry of the full theory. 

\section{Symplectic geometry of the cuscuton}\label{sec:symplectic}

Before we start, it is useful at this point to the set the notation for a smearing of a phase-space density $S$ over space: 
\begin{equation}\label{eq:smearing}
  S(N) \equiv \int \diff^3 x \, N(x) S \left[ g, \pi; x \right) \,.
\end{equation}

The Poisson bracket between arbitrary smearings of the Hamiltonian constraint of general relativity \eqref{ham-constraint-g} with itself defines the following algebra,
\begin{equation}\label{eq:constraint-algebra}
  \begin{split}
    \mathcal{A}_\text{g}(N_1,N_2) \equiv&\, \left\{\mathcal{H}_\text{g} (N_1), \mathcal{H}_\text{g} (N_2) \right\} \\
    =&\, \mathcal{H}^a_\text{g} \left( N_1\nabla_a N_2 - N_2 \nabla_a  N_1 \right) \,,
  \end{split}
\end{equation}
which corresponds to the lapse component of the algebra of deformations of general relativity in the ADM formulation. In order to determine whether the Hamiltonian constraint of the cuscuton field coupled to gravity in fact spoils refoliation invariance by selecting a preferred foliation or restricting the choice of foliations with respect to general relativity, we must determine the algebra defined by the Hamiltonian constraint of the full theory and compare it with Eq.\ \eqref{eq:constraint-algebra}.

For that purpose, we calculate the Poisson bracket between arbitrary smearings of the Hamiltonian constraint of the full theory \eqref{Ham_constraint}, that is,
\begin{equation}
  \mathcal{A}(N_1,N_2) \equiv \left\{(\mathcal{H}_\phi+\mathcal{H}_g)(N_1),  (\mathcal{H}_\phi+\mathcal{H}_g)(N_2)\right\} \,.
\end{equation}
Using definition \eqref{eq:smearing}, and dropping the dependence of $\mathcal{A}$ on the smearings, we get the following property,
\begin{equation}
  \begin{split}
    \mathcal{A} =&\, \mathcal{A}_\text{g} + \left\{\mathcal{H}_\phi(N_1), \mathcal{H}_\phi(N_2)\right\}\\
    &+ \left\{\mathcal{H}_\text{g} (N_1), \mathcal{H}_\phi (N_2) \right\} - \left\{\mathcal{H}_\text{g} (N_2), \mathcal{H}_\phi (N_1) \right\} \,.
  \end{split}
\end{equation}
The off-diagonal terms vanish, that is,
\begin{equation}\label{off_diagonal}
  \left\{\mathcal{H}_g(N_1), \mathcal{H}_\phi(N_2)\right\} - \left\{\mathcal{H}_\text{g} (N_2), \mathcal{H}_\phi (N_1) \right\} = 0 \,.
\end{equation}
This happens because there are only derivatives of the configuration variables, and ${H}_g$ has no $\pi_\phi$ dependence, and ${H}_\phi$ has no metric momenta; this implies no integration by parts needs to be done to remove derivatives from delta functions in these brackets, and thus the lapses appear without derivatives. Anti-symmetry then implies the vanishing of Eq.\ \eqref{off_diagonal}.

We now compute the final term. Remembering that $V(\phi)$ contains no derivatives of the metric, already excluding terms that are linear in $N$, and using the results from Eq.\ \eqref{eq:var-ham-const}, we obtain
\begin{equation}
  \begin{split}\label{algebra}
    \left\{\mathcal{H}_\phi(N_1), \mathcal{H}_\phi(N_2)\right\} =&\, \pi_\phi \!\nabla^a \phi \left( N_1 \nabla_a N_2 \!-\! N_2 \nabla_a N_1 \right) \\
    =&\, \mathcal{H}^a_\phi \left( N_1 \nabla_a N_2 - N_2 \nabla_a N_1 \right) \,,
  \end{split}
\end{equation}
which is of course, the same algebra as the gravitational scalar constraints obey in Eq.\ \eqref{eq:constraint-algebra}. The deformation algebra of the full Hamiltonian then is
\begin{equation}
  \mathcal{A} = \left( \mathcal{H}_\phi^a + \mathcal{H}_\text{g}^a \right) \left( N_1 \nabla_a N_2 - N_2 \nabla_a N_1 \right) \,.
\end{equation}

Thus, we obtain a consistent first class system of constraints. The number of physical degrees of freedom can be obtained from the usual counting: the dimension of the phase space $\Omega$ in a field theory in $N$-dimensional space subject to $M$ first class constraints and $S$ second class constraints is given by \cite{Henneaux:1992ig}
% %
\begin{equation}\label{eq:dofcount}
  \dim(\Omega) = \frac{1}{2} \left( N - 2 M - S \right) \,.
\end{equation}
In this case, the constraints are the same as in ADM, but we have an extra scalar field, thus we obtain a 3-dimensional space of physical degrees of freedom per space point. 

\section{The homogeneous limit}\label{sec:homog-strong}

The modulus in the square root in the cuscuton Lagrangian \eqref{cuscuton-lagrangian} ensures that the Hamiltonian \eqref{eq:cuscuton-H} remains real even if $X < 0$. However, the positive branch presents a problem on the surface $\pi_\phi = \mu^2 \sqrt{\gamma}$, as can be seen from the fact that Eq.\ \eqref{eq:dot-phi} becomes divergent there. This can be solved if the numerator vanishes at the same rate; however, it can be seen from the momentum equation of motion \eqref{eq:dot_pi_phi} that a homogeneous solution with $\nabla_i \phi = 0$ may also pose a problem, in the form of the ill-defined limit inside the divergence. Since most applications of the cuscuton involve its cosmological evolution and perturbations around a homogeneous background, in the general theory one needs to check what happens if and when the Hamiltonian flow takes the system across these surfaces. As a first step, let us analyze the behavior of the dynamical system when we impose homogeneity as a constraint.

\subsection{Weakly imposing homogeneity}

Consider the following constraint, which represents a homogeneous cuscuton field on-shell:
\begin{equation}\label{eq:homog-weak}
  \Phi_\text{h} \equiv \nabla_i \phi \approx 0 \,.
\end{equation}
Since this constraint is a pure divergence, any smearing $\zeta^i$ of Eq.\ \eqref{eq:homog-weak} must obey $\nabla_i\zeta^i\neq 0$ as other components of the smeared constraint automatically vanish. 

If one considers imposing the constraint \eqref{eq:homog-weak} onto the Hamiltonian \eqref{eq:cuscuton-H}, the equation of motion \eqref{eq:dot_pi_phi} becomes undetermined. In order to contain this potential loss of predictability, one may define the unit vector
\begin{equation}
  f^i \equiv \frac{\nabla^i \phi}{\sqrt{\nabla^a \phi \nabla_a \phi}} \,,
\end{equation}
in which case Eq.\ \eqref{eq:dot_pi_phi} may be cast in the form

\begin{equation}\label{eq:dot-pi-phi_f}
  \dot{\pi}_\phi = \nabla_i \left( N f^i \sqrt{\pi_\phi^2 - \gamma \mu^4} + \xi^i \pi_\phi \right) - N \sqrt{\gamma} \deriv{V}{\phi} \,.
\end{equation}
By setting $\nabla_i \phi = 0$, and assuming that $f^i$ remains regular in this limit, the equations of motion \eqref{eq:dot-pi-phi_f} and \eqref{eq:dot-phi} reduce to
\begin{align}
  \dot{\pi}_\phi =&\, \nabla_i \left( N f^i \sqrt{\pi_\phi^2 - \gamma \mu^4} + \xi^i \pi_\phi \right) - N \sqrt{\gamma} \deriv{V}{\phi} \,,\\
  \dot{\phi} =&\, 0 \,.
\end{align}
Notice that the right-hand side of the equation of motion for $\phi$ vanishes, effectively decoupling the field from its momentum.

Assuming there are no boundaries, we can easily check that the constraint \eqref{eq:homog-weak} is propagated:
\begin{equation}
  \begin{split}
    \left\{ \Phi_\text{h} \left( \zeta^i \right), H (N) \right\} =&\, \int \diff^3 x \sqrt{\gamma} \left[ (\zeta^i\nabla_i\phi) \frac{1}{2} \gamma^{ab} \dot\gamma_{ab} \right.\\
    &\left. \vphantom{\frac{1}{2}} - (\nabla_i\zeta^i) \dot\phi \right] \\
    \approx&\, 0 \,,
  \end{split}
\end{equation}
so the system remains on the constraint surface over the course of evolution. However, there is no equation of motion for $\phi$, and Eq.\ \eqref{eq:dot-pi-phi_f} has no predictive power due to the presence of the undetermined vector $f^i$.

A possible way around this difficulty is to consider the effect of applying the homogeneity condition in the definition of the canonical momentum in order to define an alternative constraint that might be dynamically equivalent to homogeneity, but without the singularities introduced by the constraint \eqref{eq:homog-weak}. If we consider Eq.\ \eqref{eq:pidephi}, introducing Eq.\ \eqref{eq:homog-weak} into it results in the following constraint:
\begin{equation}\label{eq:homog-weak-alt}
  \Phi_{\text{h}'} = \pi_{\phi} - \mu^2 \sqrt{\gamma} \approx 0 \,.
\end{equation}
Clearly, it commutes with itself, but it is second class  with respect to \eqref{eq:homog-weak}:
\begin{equation}\label{eq:homog-weak-alt-secondary}
    \{ \Phi_{\text{h}'}(x),  \Phi_{\text{h}}(\zeta^i) \} =-\nabla_i\zeta^i (x)\neq 0 \,.
\end{equation}
Thus the two constraints are second class. We conclude therefore that weakly imposing homogeneity either leads to divergences or to a second class set of constraints.

\subsection{Strongly imposing homogeneity}

Imposing second class constraints requires a projection onto the constraint surface. This is equivalent to using variables where  $\phi$ is a spatial constant, and $\pi_\phi$ as well.\footnote{Note, however, that in the previous system $\pi_\phi$ is a density, which is point-dependent. We would have to redefine the variables so that $\pi_\varphi\rightarrow \pi_\varphi/\sqrt{g}$ is the new momentum, and is a spatial constant.} Alternatively,  we will impose the  constraint $\nabla_i \phi = 0$ strongly,  explicitly on the Lagrangian \eqref{cuscuton-lagrangian} and show that this also results in the same system. Denoting in this particular case the constrained field by $\varphi (t)$ and its associated momentum by $\pi_{\varphi}$, we may cast the Lagrangian \eqref{cuscuton-lagrangian} as
\begin{equation}
  L_2 = \int \diff^3 x \, N \sqrt{\gamma} \left[ \mu^2 \frac{\dot{\varphi}}{N} - V(\varphi) \right]\,.
\end{equation}
Due to the spatial constancy of $\varphi$, this can be rewritten as
\begin{equation}
  L_2 = \mu^2 \dot{\varphi} \,\int \diff^3 x \, \sqrt{\gamma}- V(\varphi)\int \diff^3 x\sqrt{\gamma} N \,,
\end{equation}
so the associated momentum reads
\begin{equation}\label{eq:homog-pi}
  \pi_{\varphi} = \mu^2 \int \diff^3 x \, \sqrt{\gamma} \,.
\end{equation}
This definition introduces the following primary constraint:
\begin{equation}\label{eq:homog-const}
  \Phi_\varphi \equiv \pi_{\varphi} - \mu^2 \int \diff^3 x \, \sqrt{\gamma} \approx 0 \,.
\end{equation}
Note that now both $\varphi$ and its associated momentum $\pi_\varphi$ are spatial constants, and from the linearity of $L_2$ in $\dot\varphi$, its Hamiltonian is just the potential term. Absorbing the primary constraint into the ones we had previously found, the total Hamiltonian now reads
\begin{equation}\label{eq:hamiltonian-homog-const}
  \begin{split}
    H =&\,H_\varphi + H_\text{g} +k \Phi_\varphi  \\
    =&\, k \left( \pi_{\varphi} - \mu^2\, \int \diff^3 x \,  \sqrt{\gamma} \right) \\
    & +\int \diff^3 x \left[\, N \left( \mathcal{H}_{\text{g}} + V \sqrt{\gamma} \right) + \xi_a \mathcal{H}_{\text{g}}^a \right] \,.
  \end{split}
\end{equation}
However, the constraints composing the Hamiltonian are not first class. Indeed, we find the following expression for the algebra between the $k$ and $N$ generators:
\begin{equation}
  \left\{ \Phi_\varphi, ( \mathcal{H}_\varphi + \mathcal{H}_\text{g} ) (x) \right\} \approx \deriv{V}{\varphi} \sqrt{\gamma}(x) - \frac{\mu^2}{2} \pi (x)\,.
\end{equation}
For this to propagate, we need the left-hand side to vanish, which provides us with a CMC condition. We denote it as
\begin{equation}\label{eq:cmc}
  \Phi_\text{CMC} = \pi - \frac{2 \sqrt{\gamma}}{\mu^2} \deriv{V}{\varphi} \approx 0 \,.
\end{equation}

Now we must insert this further constraint into the system, to see how it behaves. Propagating Eq.\ \eqref{eq:cmc}, we find
\begin{equation}\label{eq:pre-lfe}
  \begin{split}
    \left\{ \Phi_\text{CMC}, H \right\} =&\, \left( \pi^{i j} - \frac{\sqrt{\gamma}}{\mu^2} \deriv{V}{\varphi} \gamma^{i j} \right) \dot{\gamma}_{i j} \\
    &+ \gamma_{i j} \dot{\pi}^{i j} - \frac{2 \sqrt{\gamma}}{\mu^2} \deriv[2]{V}{\varphi} \dot{\varphi} \,.
  \end{split}
\end{equation}
Denoting the trace-free part of the momentum as $\sigma^{ij}$, that is,
\begin{equation}\label{trace-free}
  \sigma^{i j} \equiv \pi^{i j} - \frac{1}{3} \pi \gamma^{i j} \,,
\end{equation}
and substituting the constraints into Eq.\ \eqref{eq:pre-lfe}, we find
\begin{multline}
  \left[ \frac{12}{\gamma} \sigma^{a b} \sigma_{a b} + 6 V + \left( \frac{2}{\mu^2} \deriv{V}{\varphi} \right)^2 - 12 \nabla^2 \right] N \\+ k \left( \frac{2}{\mu^2} \deriv[2]{V}{\varphi} - \frac{3}{2} \mu^2 \right) \approx 0 \,.
\end{multline}
This is a lapse-fixing equation, which together with the CMC constraint \eqref{eq:cmc} determines a gauge-fixing for the system. 

Since with our definition the spectrum of the Laplacian is strictly negative, if 
\begin{equation}\label{pos_lapse}
  12 \sigma^{a b} \sigma_{a b} + 6 V + \left( \frac{2}{\mu^2} \deriv{V}{\varphi} \right)^2>0 \,,
\end{equation}
the operator
\begin{equation}\label{eq:Delta}
  \Delta \equiv  \left[ 12 \sigma^{a b} \sigma_{a b} + 6 V + \left( \frac{2}{\mu^2} \deriv{V}{\varphi} \right)^2 - 12 \nabla^2 \right]
\end{equation}
is invertible. In that case, as long as $\left( \frac{2}{\mu^2} \deriv[2]{V}{\varphi} - \frac{3}{2} \mu^2 \right)\neq0$, we have a unique lapse for each choice of $k$, which we will call $N_0$. Nonetheless, if the left-hand side of Eq.\ \eqref{pos_lapse} is negative, $\Delta$ can have a  phase-space-dependent finite-dimensional kernel (since $\Delta$ is elliptic and we are in a compact manifold). In that case, there would be a finite-dimensional degeneracy of $N_0$.

The constrained homogeneous cuscuton second class Hamiltonian system can be written as
\begin{equation}\label{eq:hamiltonian-homog-full}
  \begin{split}
    H =&\, k_1 \Phi_\varphi + k_2 \Phi_\text{CMC} + V (\varphi)\, \int \diff^3 x  N \sqrt{\gamma} \\
    & +\int \diff^3 x \left[\, N \mathcal{H}_{\text{g}} + \xi_a \mathcal{H}_{\text{g}}^a \right] \,.
  \end{split}
\end{equation} 
This yields the foliation-fixed dynamics, which we will describe in the next section. 

\subsection{Second class system dynamics}

Since this is a second class system, to find the equations of motion in the constraint surface we have two options: either
\begin{inparaenum}[(i)]
\item we use a Dirac bracket with the inverse of the (highly non-local) operator $\Delta$ from Eq.\ \eqref{eq:Delta}, or
\item we find conjugate variables that can solve for the two second class constraints, $\mathcal{H}_g+\mathcal{H}_\varphi$ and $\Phi_\text{CMC}$.
\end{inparaenum}
Setting these variables strongly to the value of their solution, the Dirac bracket coincides with the standard Poisson one in the remaining variables.  In general, these solutions will also be highly non-local.

A convenient choice is to use the spatial conformal factor of the metric and the trace of the momentum. We take the trace of the momentum to be conformally invariant, so the transformations read
\begin{align}
  \gamma_{i j} =&\, e^{4 \omega} \bar{\gamma}_{i j} \,,\label{metric_conf}\\
  \begin{split}
    \pi^{i j} =&\, e^{-4 \omega} \bar{\pi}^{i j}\\
    =&\, e^{-4 \omega} \bar{\sigma}^{i j} + \frac{1}{3} e^{-4 \omega} \pi \bar{\gamma}^{i j} \,.\label{pi_conf}
  \end{split}
\end{align}
The best way to understand this decomposition without introducing densities into the configuration variables is to use a reference metric and a reference density, $\eta_{ij}$,\, $\sqrt{\eta}$, such that $e^{12\omega} = \nicefrac{\gamma}{\eta}$, so that $\bar\gamma=\eta$. 

Inserting this decomposition and substituting the CMC constraint \eqref{eq:cmc} into the Hamiltonian constraint, we find a Lichnerowicz--York-type equation for this system \cite{Mercati:2014ama,gourgoulhon}:
\begin{multline}
  8 e^{2 \omega} \bar{\nabla}^2 \omega - e^{\omega} \bar{\mathcal{R}} + e^{-7 \omega} \bar{\sigma}^{a b} \bar{\sigma}_{a b} \\- e^{5 \omega} \left[ \frac{1}{3 \mu^2} \left( \deriv{V}{\varphi} \right)^2 + V \right] = 0 \,.
\end{multline}
According to the Lichnerowicz--York method, this will have unique solutions as long as $ 6 V + \left( \frac{2}{\mu^2} \deriv{V}{\varphi} \right)^2>0$, which is also a sufficient condition for the positivity of Eq.\ \eqref{pos_lapse}. This is an interesting substitution of the standard condition for the unique solutions of the LY equation, which deserves to be further explored.  

Under \eqref{trace-free},  the smeared momentum constraint reads
\begin{equation}
  \begin{split}
    \int\! \diff^3 x\, \pi^{ij}\mathcal{L}_\xi \gamma_{ij} &\rightarrow \!\int\! \diff^3x \left( \! \sigma^{ij }\mathcal{L}_\xi \gamma_{ij} + \frac{1}{3} \pi \gamma^{ij} \mathcal{L}_\xi \gamma_{ij} \right) \\
    &= \int \diff^3 x \left( \sigma^{ij} \mathcal{L}_\xi \gamma_{ij} + \pi \nabla^k \xi_k \right) \\
    &\hat{=} \int \diff^3 x \left( \sigma^{ij} \mathcal{L}_\xi \gamma_{ij} \right) \,,
\end{split}
\end{equation}
where the last term in the middle equation vanishes for when $\frac{\pi}{\sqrt{g}}$ is a spatial constant, as $\nabla^k\xi_k= \gamma^{ij}\mathcal{L}_\xi \gamma_{ij}$ is a pure divergence. Under the conformal transformations induced by Eqs.\ \eqref{metric_conf} and \eqref{pi_conf}, we now get
\begin{equation}
\int d^3x \left(\sigma^{ij}\mathcal{L}_\xi \gamma_{ij}\right)\rightarrow \int d^3x \left(\bar\sigma^{ij}\mathcal{L}_\xi \bar\gamma_{ij}\right) \,,
\end{equation}
since the extra term containing the derivative of $\omega$ is multiplied by $\bar\gamma^{ij}\sigma_{ij}$ and thus vanishes. This decouples the momentum and scalar constraints. 

Finally, the complete system of equations of motion then reads
\begin{align}
  \dot{\varphi} =&\, k \,,\\
  \dot{\pi}_\varphi =&\, - \deriv{V}{\varphi} \int \diff^3 x\, N_0 e^{6 \omega} \,,\\
  \dot{\omega} - \Lie_\xi \omega =&\, \frac{1}{6} \left( \bar{\nabla}_a \xi^a - \frac{N_0}{\mu^2} \deriv{V}{\varphi} \right) \,,\\
  \dot{\bar{\gamma}}_{i j} - \Lie_\xi \bar{\gamma}_{i j} =&\, 2 N_0 e^{-6 \omega} \bar{\sigma}_{i j} - \frac{2}{3} \bar{\nabla}_a \xi^a \bar{\gamma}_{i j} \,,\\
  \begin{split}
    \dot{\bar{\sigma}}_{i j} - \Lie{\bar{\sigma}}_{i j} =&\, -\frac{1}{3} \bar{\nabla}_a \xi^a \bar{\sigma}_{i j} + 2 N_0 e^{-2 \omega} \bar{\gamma}^{a b} \bar{\sigma}_{i a} \bar{\sigma}_{j b} \\
    & + e^{-2 \omega} \left\{\vphantom{\frac{1}{3}} \bar{\nabla}_i \bar{\nabla}_j N_0 - 4 \bar{\nabla}_{\left( i \right.} \omega \bar{\nabla}_{\left. j \right)} N_0 \right.\\
    & - \frac{1}{3} \left( \bar{\nabla}^2 N_0 - 4 \bar{\nabla}_a \omega \bar{\nabla}^a N_0 \right) \bar{\gamma}_{i j}\\
    & - N_0 \left[ \bar{\mathcal{R}}_{i j} - \frac{1}{3} \bar{\mathcal{R}} \bar{\gamma}_{i j} \right.\\
    & - 2 \bar{\nabla}_i \bar{\nabla}_j \omega + 4 \bar{\nabla}_i \omega \bar{\nabla}_j \omega \\
    & \left.\left. + \frac{2}{3} \left( \bar{\nabla}^2 \omega - 2 \bar{\nabla}_a \omega \bar{\nabla}^a \omega \right) \bar{\gamma}_{i j} \right] \right\} \,.
  \end{split}
\end{align}

Note that no source term appears in the evolution equation for $\bar{\sigma}_{i j}$. This can also be interpreted as a consequence of the energy-momentum tensor associated with the cuscuton being that of a perfect fluid (like any $k$-essence), and therefore possessing only terms that are proportional to the metric.

\section{Conclusions}\label{sec:conclusion}

In this work we have formulated the cuscuton field, an extreme form of $k$-essence, in the Hamiltonian formalism, in order to count the degrees of freedom propagated by the field. By computing the algebra of constraints in the full system, we have found no evidence that a general field requires a preferred foliation. However, the field equations become singular in the homogeneous limit, so a transition to or from a homogeneous regime implies the loss of predictivity of the equations of motion. We have also shown that this problem remains even if homogeneity is imposed as a constraint on the system, as the system becomes second class. A way to resolve this difficulty is to impose strong homogeneity of the cuscuton field, by requiring that the dynamics be described in a foliation in which no dependence of the field with respect to spatial variables appears in the Hamiltonian. We then show that the resulting theory is constrained to a CMC foliation whose propagation is ensured by a lapse-fixing equation, which effectively fixes the gauge for this theory and ensures that it is well defined. In this gauge, there are in fact no dynamical evolution equations for the cuscuton, and the time derivative of the field appears as a Lagrange multiplier which parametrizes the patch of lapse functions that solve the CMC equation. No local excitations of the field can be activated, and therefore no worries about superluminal propagation need ensue. 

Because the homogeneous limit is fundamentally discontinuous from the generic cuscuton action, our analysis shows that the two limits must be treated as different dynamical systems. Therefore, if one performs perturbation analysis around a homogeneous cuscuton background, one must carry out the analysis within the limits of a CMC foliation, and without allowing for local perturbations on the cuscuton field itself. By allowing the cuscuton to depart from homogeneity, it effectively becomes an entirely different system with nothing of the constraint structure that appears in the homogeneous case. In this latter case, it acquires an extra locally propagating scalar degree of freedom.

Regarding the inhomogeneous case, the presence of two singular surfaces in phase space which are not limit surfaces --- and which may in principle be reached in finite time by the Hamiltonian flow of a generic initial condition --- may in fact render the inhomogeneous cuscuton a non-physical model, in the sense that any generic initial condition surface may lead to an ill-defined system. However, a full analysis of the dynamical evolution of the inhomogeneous cuscuton is beyond the scope of this work, and will be the subject of future study.

The homogeneous cuscuton and the structure it generates has an interesting consequence. The CMC condition \eqref{eq:cmc} establishes a connection between the trace of the gravitational momentum and the cuscuton potential. This means that each choice of solution for the potential corresponds to a different function of the momentum to appear in the Hamiltonian \cite{Afshordi:2010eq}. Gravitational theories which propose a dependency on the trace of the gravitational momentum include time-asymmetric modifications of general relativity \cite{Gomes:2013naa,*Cortes:2015ola} and the linking theory connecting shape dynamics and general relativity \cite{deA.Gomes:2011wk,*Gomes:2011zi,Mercati:2014ama}. If these theories can be shown to display the same constraint structure as the cuscuton for a suitable choice of potentials, this in turn means that each cuscuton potential corresponds to an Einstein-frame description of a modified gravity theory with some dependence on the trace of the gravitational momentum.\footnote{In the case of shape dynamics, this will only be a correspondence over a given patch of the conformal phase space which can be covered by the ADM gauge-section.}

The ubiquity of CMC foliations in many different attempts to implement in GR a ``time-function'' --- \emph{i.e.} a clock synchronization which is integrable and characterized by the constant value of some scalar field --- raises questions: could it be related to the role that such foliations play in allowing a duality with spatially conformally invariant theories \cite{Gomes:2010fh,Gomes:2011zi,Gomes:2013naa}? Or it could be related to the fact that, up to second order in gravitational momenta, there are no first class extensions of the ADM scalar constraint \cite{Gomes:2016vqk}, and $\pi= \text{constant}$ is the only spatially-covariant, first class constraint, which is purely second class with respect to $\mathcal{H}_g$ (given in Eq.\ \eqref{ham-constraint-g}) over the entire ADM phase space \cite{Gomes:2013naa}. Of course, these two reasons are not mutually exclusive, and have in fact been argued to be related \cite{Gomes:2013naa}.  

Lastly, we comment on the substitution of York time by the appearance of $ 6 V + \left( \frac{2}{\mu^2} \deriv{V}{\varphi} \right)^2$.   In the context of cosmology, York time is the Hubble parameter $H$ up to a negative constant,\footnote{For a more thorough study of cosmology in York time, see Ref.\ \cite{ellis-cosmology}.}
\begin{equation}\label{eq.THincosm}
  \tau = -(4\pi G)^{-1} H = -2M_{\text{Pl}}^{2} H \,,
\end{equation}
where $G$ is the gravitational constant and $M_{\text{Pl}}$ the reduced Planck mass.  The monotonicity of $\tau$, a requirement for it to function as a time parameter, is guaranteed if the equation of state parameter $w$ obeys $w\geq-1$, %This is a result of substituting the mass-energy conservation equation with energy density $\rho$ and pressure $p$,
%\begin{equation} \dot{\rho} = -3H(\rho+p), \end{equation}
%into the time derivative of the first Friedmann equation,
%\begin{equation} 2H\dot{H} = (3M_{Pl}^2)^{-1}\dot{\rho}.\end{equation}
%A dot denotes a derivative with respect to $t$. All conventional matter content obeys $w>-1$. For instance, in the case of a universe dominated by a scalar field one has 
%\begin{equation}\rho=\frac12\dot{\phi}^2+V(\phi),\qquad p=\frac12\dot{\phi}^2-V(\phi),\end{equation}which implies $w\geq-1$
 with equality only in the limiting case when the kinetic contribution vanishes exactly, a scenario corresponding to a first approximation to slow-roll inflation. Here, we could potentially adjust the potential $V(\varphi)$ to probe different domains, making it more flexible than York time. These questions remain for further study. 
  
\begin{acknowledgments}

We thank N.\ Afshordi, G.\ Geshnizjani, T.\ Koslowski, F.\ Mercati, L.\ Smolin and R.\ Sorkin for insights and valuable discussions. This research was supported in part by Perimeter Institute for Theoretical Physics. Research at Perimeter Institute is supported by the Government of Canada through Innovation, Science and Economic Development Canada and by the Province of Ontario through the Ministry of Research, Innovation and Science.

\end{acknowledgments}

\appendix

\section{A quick primer on symplectic geometry}

In the Hamiltonian formalism, the symplectic flow of constraints generates motion in phase space. Because the Poisson bracket acts as a derivative,
\begin{equation}
  \left\{ f, g h \right\} = g \left\{ f, h \right\} + \left\{ f, g \right\} h \,,
\end{equation}
we can see the linear operator $\left\{ f, \cdot \right\}$ as a kind of vector field in phase space. Thus, the \emph{symplectic flow} of a given phase-space function $f$ is defined as $v_f \equiv \left\{ f, \cdot \right\}$. It will act on other functions as a directional derivative ${v}_f[h]$ and measure how much $h$ changes in the direction of ${v}_f$. That is, it measures how other phase-space functions change  under ``evolution'' through the action of the corresponding phase-space function $f$. The phase-space function $f$ implicitly defines a surface through the regular value theorem, provided certain regularity assumptions are satisfied. In the presence of a metric, one would usually say that the differential one-form $\diff f$ is ``perpendicular'' to the surface $f^{-1}(0)$, because its dual vector field $\diff f^\sharp$ is defined as $X[f]=\diff f(X) \eqqcolon g(\diff f^\sharp,X)$, \emph{i.e.} $g^{-1}\diff f=(\diff f)^\sharp$, which obviously vanishes for  any vector field tangent to $f^{-1}(0)$. In the case of symplectic geometry, one does not define the analogous operation through the use of a metric but a symplectic two-form, usually denoted by $\omega$. Explicitly,
\begin{equation}\label{equ:symplectic_form1}
  \omega({v}_f,\cdot) \coloneqq \diff f \,,
\end{equation}
and furthermore
\begin{equation}\label{equ:symplectic_form2}
  \omega({v}_f,{v}_h)=\{f, h\}.
\end{equation}

Now, just as a vector field can be tangential to a given manifold, so can symplectic flows. Suppose then that a surface $\mathcal{N}$ in phase space is given by the intersection of regular manifolds defined by the inverse values of the functions $\chi^I$, \emph{i.e.}, $\mathcal{N} \coloneqq \{(q,p)~|~\chi^I(q,p)=0 ~\forall I \}$.  Then $\mathcal{N}$ will be said to be \emph{first class} if: for all phase-space functions $f$ such that $f$ vanishes on $\mathcal{N}$, \emph{i.e.} $\diff f(X)=0$ for all $X \in T \mathcal{N}$, then ${v}_f [\chi^I] (p, q) = 0$ for all $I$ and $(p, q) \in \mathcal{N}$. The statement is equivalent to the much simpler statement that $\{f, \chi^I \} = a_J \chi^J$, since this will indeed be zero whenever we are on the surface. The geometric translation is indeed very simple: $\mathcal{N}$ is first class if all symplectic flows ${v}_f$ of functions $f$ that vanish on the surface $\mathcal{N}$ are tangent to the surface.

By contrast, we can define a second class manifold (or set of regular functions $\chi^I$) if all  symplectic flows (of functions that vanish on the surface) take us out of the surface (\emph{i.e.} are not tangent to it).

Gauge-fixings are further constraints imposed to be second class with the symmetry generator one would like to fix. Its flow is transverse to the gauge-orbits, and it does not change the physical degree of freedom count (a theory with one set of $n$ first class constraints has the same number of degrees of freedom than a theory with $2n$ second class constraints).  On the other hand, imposition of a further first class constraint will change the degree of freedom counting, consisting in an actual dynamical reduction. 

\section{Equations of motion in the $X < 0$ region}\label{sec:Xneg}

It can immediately be seen from the definition of $X$ in Eq.\ \eqref{Xdef} that the region $X < 0$ does not intersect with the homogeneous region. However, it is a simple exercise to demonstrate that this region of the cuscuton is well-behaved, at least in the sense of the constraint algebra.

Let us consider $X < 0$ in the Lagrangian \eqref{cuscuton-lagrangian}. The conjugate momentum in this case reads
\begin{equation}
  \pi_\phi = -\frac{\sqrt{\gamma}}{N} \frac{\mu^2}{\sqrt{-2 X}} \left( \dot{\phi} - \xi^i \nabla_i \phi \right) \,,
\end{equation}
so when we cast $\dot{\phi}$ and $X$ in terms of the momentum we find
\begin{align}
  \dot{\phi} =&\, \xi^i \nabla_i \phi - N \pi_\phi \sqrt{\frac{\nabla^i \phi \nabla_i \phi}{\pi_\phi^2 + \gamma \mu^4}} \,,\\
  X =&\, -\frac{1}{2} \nabla^i \phi \nabla_i \phi \left( \frac{\gamma \mu^4}{\pi_\phi^2 + \gamma \mu^4} \right) \,,
\end{align}
The Hamiltonian is then given by
\begin{equation}
  \begin{split}
    H_\phi =& \int \diff^ 3x\,\left[ N \sqrt{\gamma} V - N \sqrt{\nabla^i \phi \nabla_i \phi \left( \pi_\phi^2 + \gamma \mu^4 \right)} \right.\\
    &\left. \vphantom{\sqrt{\nabla^i \phi \nabla_i \phi \left( \pi_\phi^2 - \gamma \mu^4 \right)}} + \xi_i \pi_\phi \nabla^i \phi \right]\,,
  \end{split}
\end{equation}
and the equations of motion read
\begin{align}
  \begin{split}
    \dot{\pi}_\phi =& - \nabla_i \left( N \nabla^i \phi \sqrt{\frac{\pi_\phi^2 + \gamma \mu^4}{\nabla^a \phi \nabla_a \phi}} - \xi^i \pi_\phi \right) \\
    &- N \sqrt{\gamma} \deriv{V}{\phi} \,,
  \end{split}\\
  \dot{\phi} =&\, \xi^i \nabla_i \phi - N \pi_\phi \sqrt{\frac{\nabla^i \phi \nabla_i \phi}{\pi_\phi^2 + \gamma \mu^4}} \,.
\end{align}
Including the gravitational field, the full Hamiltonian constraint reads
\begin{equation}
  \begin{split}
    \left\{ H, \pi_N \right\} =&\, \frac{1}{\sqrt{\gamma}} \left( \pi^{a b} \pi_{a b} - \frac{1}{2} \pi^2 \right) - \sqrt{\gamma} \mathcal{R} \\
    & - \sqrt{\nabla^i \phi \nabla_i \phi \left( \pi_\phi^2 + \gamma \mu^4 \right)} + \sqrt{\gamma} V \\
    =&\, 0 \,,
  \end{split}
\end{equation}
while the momentum constraint remains unaltered. From these results, it becomes clear that the negative region obeys the same deformation algebra \eqref{algebra} as the positive region.

\section{Full variation of the cuscuton Hamiltonian}\label{sec:full-var}

In this Appendix we present all components of the variation of the cuscuton Hamiltonian. The components of the variation of the supermomentum term $\mathcal{H}^i_\phi$ with respect to the field variables are
\begin{widetext}
\begin{equation}
  \begin{split}
    \int \diff^3 x \, \delta \left( \xi^i \mathcal{H}_i^\phi \right) =& \int \diff^3 x \, \left[ \pi_\phi \nabla_i \phi \, \delta \xi^i + \xi^i \delta \left( \pi_\phi \nabla_i \phi \right) \right]\\
    =& \int \diff^3 x \left[ \pi_\phi \nabla_i \phi \, \delta \xi^i + \xi^i \nabla_i \phi \delta \pi_\phi + \pi_\phi \xi^i \nabla_i \left(\delta \phi \right) \right]\\
    =& \int \diff^3 x \left[ \pi_\phi \nabla_i \phi \, \delta \xi^i + \xi^i \nabla_i \phi \delta \pi_\phi - \nabla_i \left( \xi^i \pi_\phi \right) \delta \phi + \nabla_i \left( \pi_\phi \xi^i \delta \phi \right) \right]\\
    =& \int \diff^3 x \left[ \pi_\phi \nabla_i \phi \, \delta \xi^i + \xi^i \nabla_i \phi \delta \pi_\phi - \nabla_i \left( \xi^i \pi_\phi \right) \delta \phi \right] + \oint \diff^2 s_i \pi_\phi \xi^i \delta \phi \,.\\
  \end{split}
\end{equation}
Likewise, the variation of the superhamiltonian term $\mathcal{H}_\phi$ reads
\begin{equation}\label{eq:var-ham-const}
  \begin{split}
    \int \diff^3 x \delta \left( N \mathcal{H}_\phi \right) =& \int \diff^3 x \, \left\{ \mathcal{H}_\phi \delta N + N \delta \left[ \sqrt{\nabla^j \phi \nabla_j \phi \left( \pi_\phi^2 - \gamma \mu^4 \right)} + \sqrt{\gamma} V \right] \right\}\\
=& \int \diff^3 x \, \left\{ \mathcal{H}_\phi \delta N + N \left[ \fderiv{\mathcal{H}_\phi}{\phi} \delta \phi + \fderiv{\mathcal{H}_\phi}{(\nabla \phi)} \delta \left( \nabla \phi \right) + \fderiv{\mathcal{H}_\phi}{\pi_\phi} \delta \pi_\phi + \left( \fderiv{\mathcal{H}_\phi}{{\gamma^{i j}}} - \fderiv{\mathcal{H}_\phi}{\gamma} \gamma \gamma_{i j}  \right) \delta \gamma^{i j} \right] \right\}\\
    =& \int \diff^3 x \, \left\{ \mathcal{H}_\phi \delta N + N \sqrt{\gamma} \deriv{V}{\phi} \delta\phi + N \sqrt{\frac{\pi_\phi^2 - \gamma \mu^4}{\nabla^j \phi \nabla_j \phi}} \nabla^i \phi \nabla_i \left(\delta \phi \right) \right.\\
    & \left. + N \pi_\phi \sqrt{\frac{\nabla^j \phi \nabla_j \phi}{\pi_\phi^2 - \gamma \mu^4}} \delta \pi + N \left( \fderiv{\mathcal{H}_\phi}{{\gamma^{i j}}} - \fderiv{\mathcal{H}_\phi}{\gamma} \gamma \gamma_{i j}  \right) \delta \gamma^{i j} \right\}\\
    =& \int \diff^3 x \left\{ \left[ \sqrt{\nabla^a \phi \nabla_a \phi \left( \pi_\phi^2 - \gamma \mu^4 \right)} + \sqrt{\gamma} V \right] \delta N \right.\\
    &  + \left[ N \sqrt{\gamma} \deriv{V}{\phi} - \nabla_i \left( N \nabla^i \phi \sqrt{\frac{\pi_\phi^2 - \gamma \mu^4}{\nabla^a \phi \nabla_a \phi}} \right) \right] \delta\phi + N \pi_\phi \sqrt{\frac{\nabla^a \phi \nabla_a \phi}{\pi_\phi^2 - \gamma \mu^4}} \delta \pi \\
      & \left. + \frac{N}{2} \left[ \nabla_i \phi \nabla_j \phi \sqrt{\frac{\pi_\phi^2 - \gamma \mu^4}{\nabla^a \phi \nabla_a \phi}} + \gamma_{i j} \left( \gamma \mu^4 \sqrt{\frac{\nabla^a \phi \nabla_a \phi}{\pi_\phi^2 - \gamma \mu^4}} - V \sqrt{\gamma} \right) \right] \delta \gamma^{i j} \right\} \\
    & + \oint \diff^2 s_i \, N \nabla^i \phi \sqrt{\frac{\pi_\phi^2 - \gamma \mu^4}{\nabla^a \phi \nabla_a \phi}} \delta \phi \,.
  \end{split}
\end{equation}
\end{widetext}

\bibliography{shortnames,referencias}

\end{document}